\documentclass[iop]{emulateapj}                                            %

\newcommand{\be}{\begin{equation}}
\newcommand{\ee}{\end{equation}}
\newcommand{\ba}{\begin{eqnarray}}
\newcommand{\ea}{\end{eqnarray}}

\newcommand{\saxj}{\mbox{SAX~J1808.4-3658 }}

\tabletypesize{\small}

\newcommand{\km}{\hbox{km}}

\slugcomment{To Appear in the Astrophysical Journal}
\shorttitle{Multi-epoch Analysis of SAX J1808.4-3658}

\begin{document}
\title{Multi-epoch Analysis of Pulse Shapes from the Neutron Star SAX~J1808.4-3658}
\author {Sharon M. Morsink\altaffilmark{1} \& Denis A. Leahy\altaffilmark{2}
}
\email{morsink@ualberta.ca, leahy@ucalgary.ca }

\altaffiltext{1}{Department of Physics,
University of Alberta, Edmonton, AB, T6G~2G7, Canada}
\altaffiltext{2}{Department of Physics and Astronomy, University of Calgary,
Calgary AB, T2N~1N4, Canada}

\begin{abstract}

The pulse shapes detected during multiple outbursts of \saxj are analyzed in
order to constrain the neutron star's mass and radius. We use a hot-spot model
with a small scattered-light component
to jointly fit data from two different epochs, under the restriction that the
star's mass and radius and the binary's inclination do not change from epoch to 
epoch. All other parameters describing the spot location, emissivity, 
and relative fractions of blackbody to Comptonized radiation are allowed to 
vary with time. The joint fit of data from the 1998 ``slow decay'' and 
the 2002 ``end of outburst maximum'' epochs using the constraint $i<90^\circ$
leads to the $3\sigma$ confidence constraint on the neutron star mass 
$ 0.8 M_\odot < M < 1.7 M_\odot$
and equatorial radius $5 \km < R < 13 \km$.  Inclinations as low as $41^\circ$ are allowed.  
The best-fit models with $M > 1.0 M_\odot$ from joint fits of the 1998 data with
data from other epochs of the 2002 and 2005 outbursts
also fall within the same $3\sigma$ confidence region. This $3\sigma$ confidence region
allows a wide variety of hadronic equations of state, in contrast with an earlier 
analysis \citep{LMC08} of only the 1998 outburst data that only allowed for extremely small stars.

\end{abstract}


\section{Introduction}
\label{s:intro}

The discovery \citep{WvdK98} of the first accretion powered
millisecond pulsar, SAX J1808.4 3658 (hereafter SAX J1808)),
confirms the proposal that neutron stars in low-mass X-ray binaries
are the progenitors of millisecond-period pulsars. SAX J1808
is now one of eleven known accreting millisecond X-ray pulsars
(see \cite{W06} and \cite{P06} for reviews).
The X-ray pulsations are most likely
produced from accretion onto the neutron star's magnetic poles (see, e.g., Fig. 12
of \citet{GDB02}). Spectral models \citep{GDB02} provide strong evidence that the X-rays
correspond to blackbody emission from a spot on the star which
is then Compton scattered by electrons above the hot spot. 
Since the pulsed light is emitted from (or very close to) the neutron star's
surface, accreting
millisecond X-ray pulsars are excellent targets for light-curve fitting
in order to constrain the neutron star equation of state (EOS).
The X-ray light curve depends on the intrinsic properties of the
emission spot (size, shape, location, and emissivity), as well as the
neutron star's properties (mass, radius, and spin).
Constraints on the star's mass and radius lead to constraints on the EOS of supernuclear density material.

The first pulse shape analysis (\citet{PG03},
hereafter PG03) for SAX J1808 provided interesting constraints
on the neutron star's mass and radius. However, this analysis did
not take into account two effects that are important for
rapidly rotating neutron stars: variable time delays due to light
travel time across the star \citep{CLM05} and the oblate
shape of the star \citep{CMLC07}. The pulse shape analysis including
these effects was done by \cite{LMC08}, resulting in smaller mass
and radius than found in PG03. Since the pulse shapes of
SAX J1808 were variable during the time period analyzed
by PG03 and during subsequent outbursts by SAX J1808 (see the comprehensive analysis by \citet{Har08}), 
it is important include the time variability in the analysis. In effect,
the results of an analysis of a time-averaged pulse profile and an analysis
of individually different pulse profiles are expected to be different.
Here we carry out an analysis of a set of different pulse profiles from different time periods from SAX J1808. 
In addition, fitting multiple pulse profiles with the same model should provide a much more stringent test
 of the validity of the model than fitting of a single pulse profile. 

The outline of our paper is as follows. In section 2, we introduce the
different pulse shapes observed from SAX J1808, which come from different phases of the 1998, 2002 and 2005 outbursts.
In section 3, the hot spot model is described and we note that an additional
scattered-light component is required in order to fit all of the pulse shapes.
 In section 4, the results are given for joint fits of the pulse shape data, including resulting mass, 
radius and inclination values. We conclude
with a general discussion of the results, comparing with previous work.

\section{Data}
\label{s:data}

SAX J1808 was observed by RXTE during the 1998, 2002, 2005 and 2008 outbursts. In this paper we concentrate
on the publicly available data from 1998, 2002 and 2005. Since the pulse shapes detected during the 2008 
outburst are very similar to the pulse shapes detected in the earlier outbursts \citep{Har09}, the addition of the
2008 data is not expected to add significant information to the analysis. 
\citet{Har08} showed that the pulse shape is variable during 
an outburst and found shorter time periods over which the pulse shape is stable, as shown in Figure 3 of 
their paper. The notation introduced by \citet{Har08} for each time period of an outburst is: 1, burst rise;
2, beginning of outburst maximum; 3, end of the maximum; 4, slow decay; 5, steep luminosity drop; and 6, 
the flaring tail. We make use of the same time intervals and notation presented by \citet{Har08}. For instance,
1998B4 refers to the 1998 outburst during slow decay stage, shown in Box 4 of Figure 3 in \citet{Har08}. 
In our analysis, we can only make use of data that has a high signal-to-noise ratio so that meaningful fits 
to the data can be performed. This results in a total of 7 time intervals, which are used in this 
paper: 
1998B4, 2002B3, 2002B4, 2005B1, 2005B2, 2005B3, and 2005B4. 
In the case of 2005B4 the data from 2005B4A and 2005B4B were combined 
in order to provide a stronger signal. The flux from the source is very low during stages 5 and 6 so 
unfortunately useful pulse shapes are not available. In all cases, the pulse shapes are binned to 32 bins
per pulse period.

The data for each time interval are binned into two narrow energy bands, a low-energy band from 3 - 4 keV
and a high-energy band from 9 - 20 keV. This choice of energy bands is motivated by previous spectral models
\citep{GDB02,PG03} for the pulsed emission that includes two components: blackbody emission and a 
Comptonized powerlaw. The high-energy band is dominated by the powerlaw component, while the blackbody component
makes a significant contribution to the low-energy band. This choice of bands avoids the energy region
between 5 and 8 keV which includes the iron line contribution from the disk \citep{Cac09,Pap09}. 

\begin{figure}
\plotone{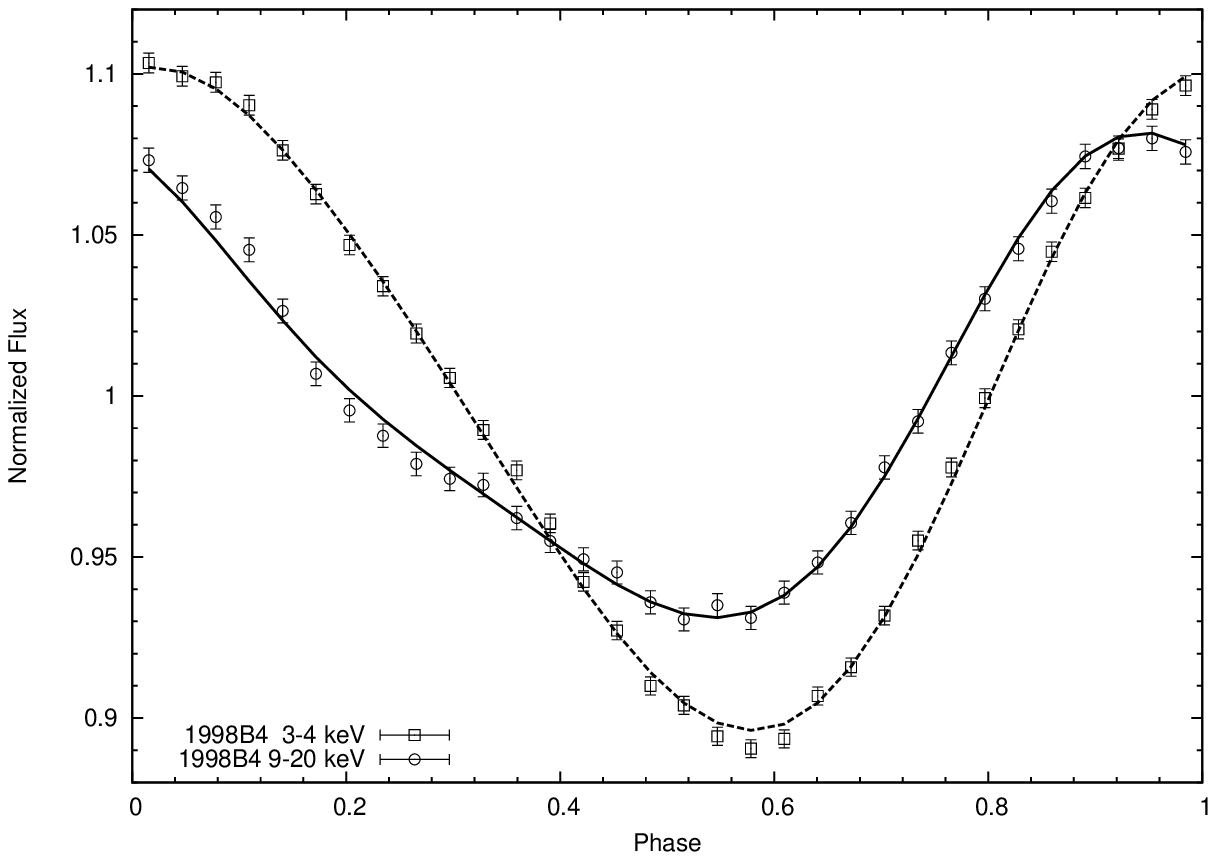}
\caption{Data from the 1998B4 period of the 1998 outburst (squares 3-4 keV, circles 9-20 keV). The solid 
and dashed curves correspond to the best-fit model that results from
a joint fit of the 1998B4 and 2002B3 data sets. This model (see row 1 of
Table \ref{tab:joint98-02B3}) has $2M/R=0.6$, $M=1.31 M_\odot$ and 
$\chi^2 = 116$ for 110 degrees of freedom.
}
\label{fig:98b4}
\end{figure}

\begin{figure}
\plotone{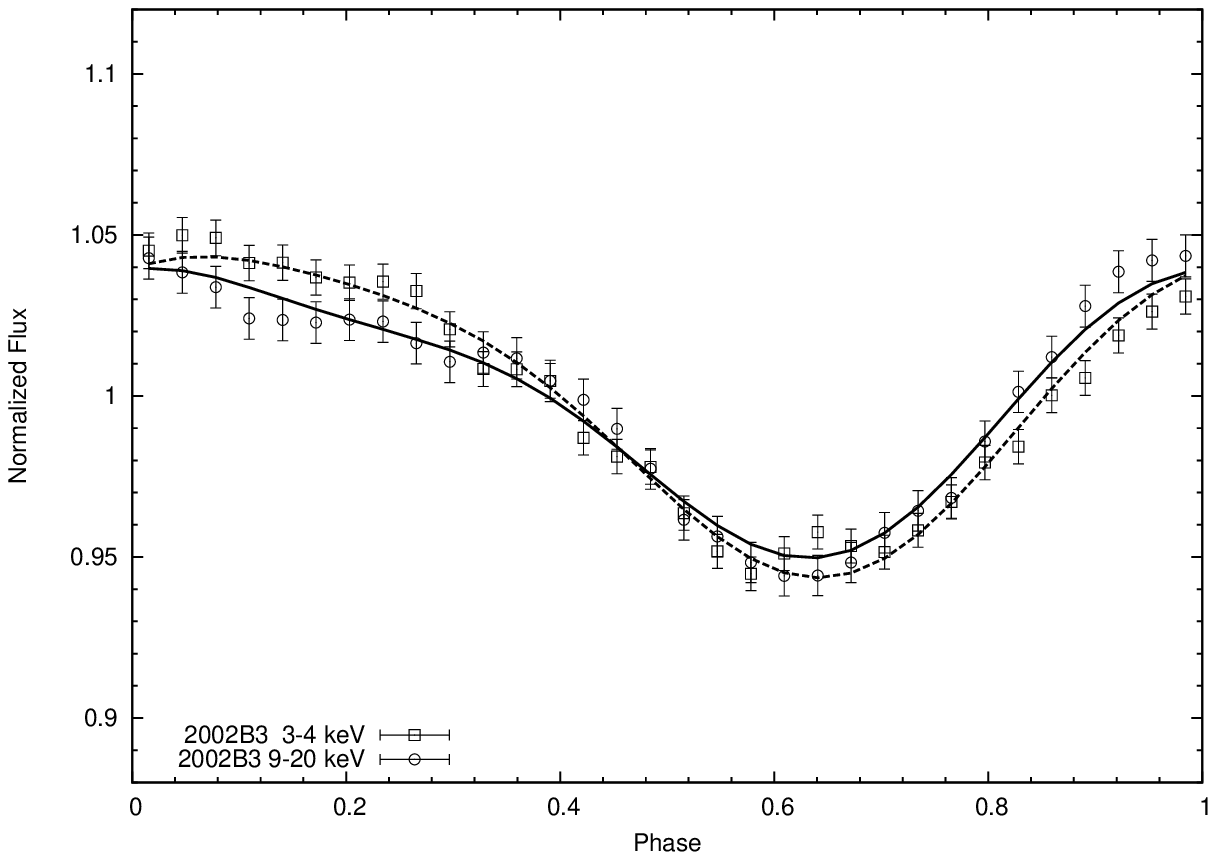}
\caption{Data from the 2002B3 period of the 2002 outburst (squares 3-4 keV, circles 9-20 keV). The solid 
and dashed curves correspond to the best-fit model that results from
a joint fit of the 1998B4 and 2002B3 data sets. This model (see row 1 of
Table \ref{tab:joint98-02B3}) has $2M/R=0.6$, $M=1.31 M_\odot$ and 
$\chi^2 = 116$ for 110 degrees of freedom.
}
\label{fig:02b3}
\end{figure}

\begin{figure}
\plotone{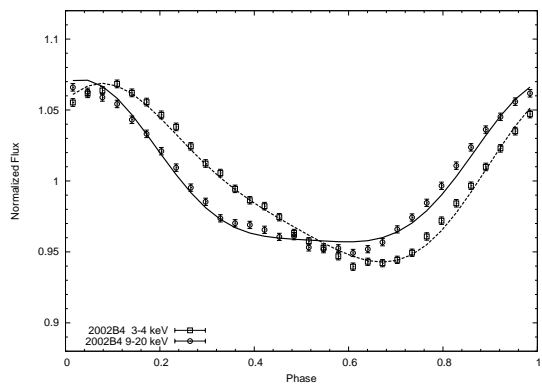}
\caption{Data from the 2002B4 period of the 2002 outburst (squares 3-4 keV, circles 9-20 keV). The solid 
and dashed curves correspond to the best-fit model that results from
a joint fit of the 1998B4 and 2002B4 data sets. This model (see row 1 of
Table \ref{tab:joint98-02B4}) has $2M/R=0.6$, $M=1.27 M_\odot$ and 
$\chi^2 = 106$ for 110 degrees of freedom. The best-fit light curves 
for the 1998B4 data are not plotted since the curves are not easily
distinguishable from the curves plotted in Figure~\ref{fig:98b4}.
}
\label{fig:02b4}
\end{figure}

\begin{figure}
\plotone{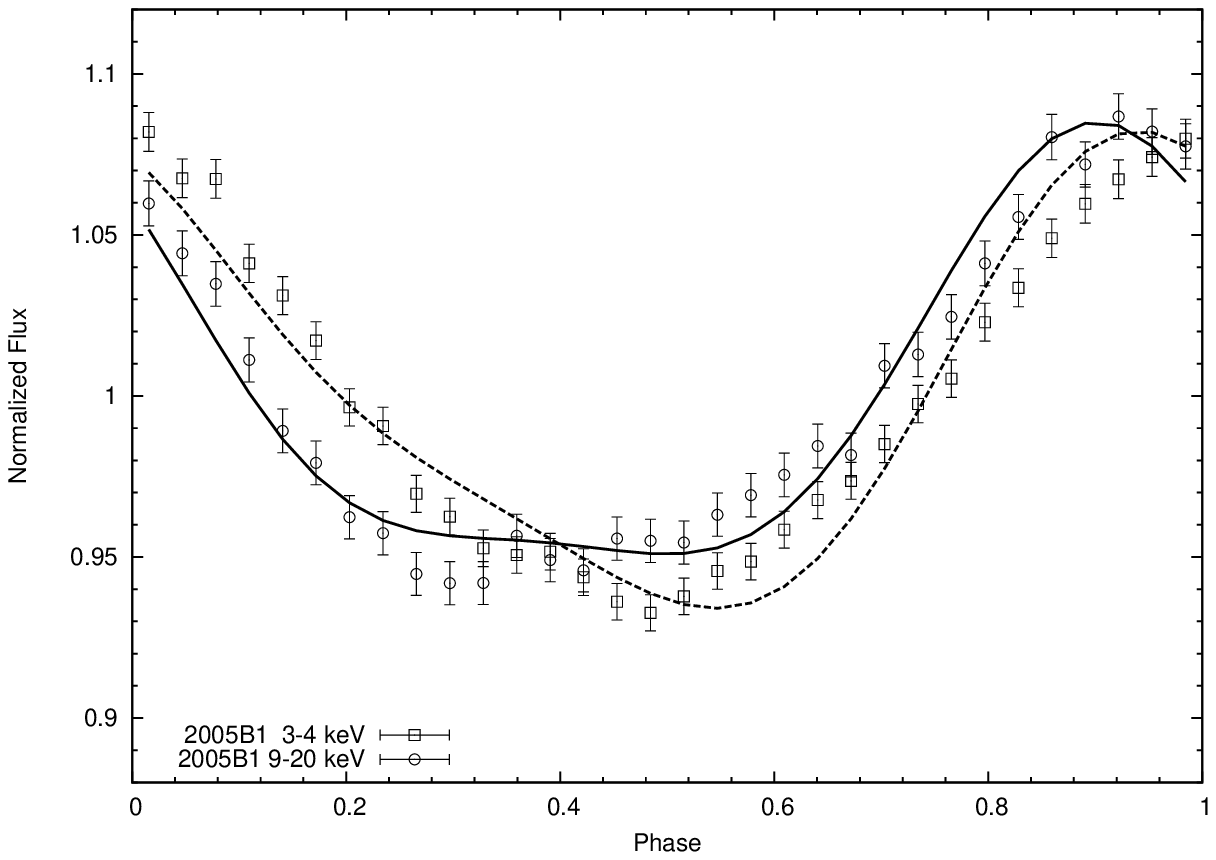}
\caption{Data from the 2005B1 period of the 2005 outburst (squares 3-4 keV, circles 9-20 keV). The solid 
and dashed curves correspond to the best-fit model that results from
a joint fit of the 1998B4 and 2005B1 data sets. This model (see row 1 of
Table \ref{tab:joint98-05B1}) has $2M/R=0.6$, $M=1.36 M_\odot$ and 
$\chi^2 = 111$ for 110 degrees of freedom.
}
\label{fig:05b1}
\end{figure}

\begin{figure}
\plotone{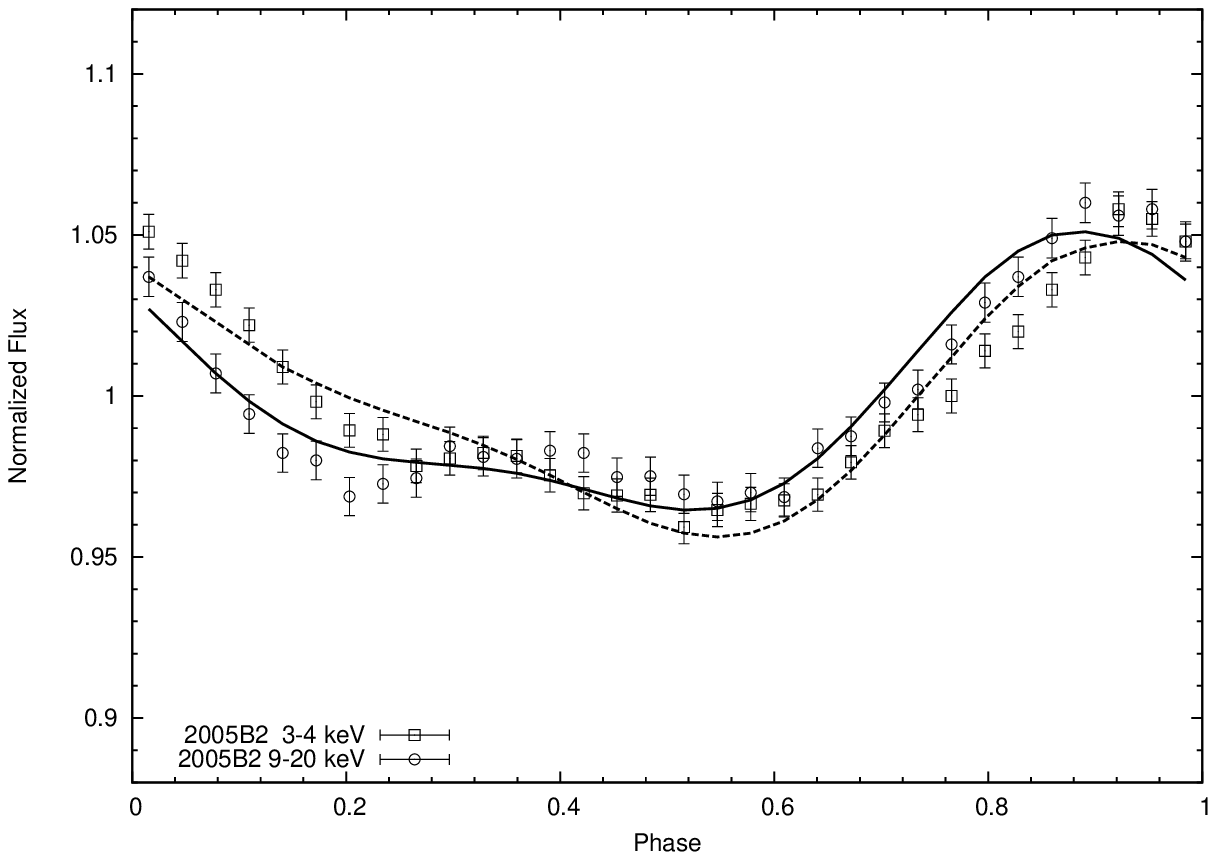}
\caption{Data from the 2005B2 period of the 2005 outburst (squares 3-4 keV, circles 9-20 keV). The solid 
and dashed curves correspond to the best-fit model that results from
a joint fit of the 1998B4 and 2005B2 data sets. This model (see row 1 of
Table \ref{tab:joint98-05B2}) has $2M/R=0.6$, $M=1.20 M_\odot$ and 
$\chi^2 = 109$ for 110 degrees of freedom.
}
\label{fig:05b2}
\end{figure}

\begin{figure}
\plotone{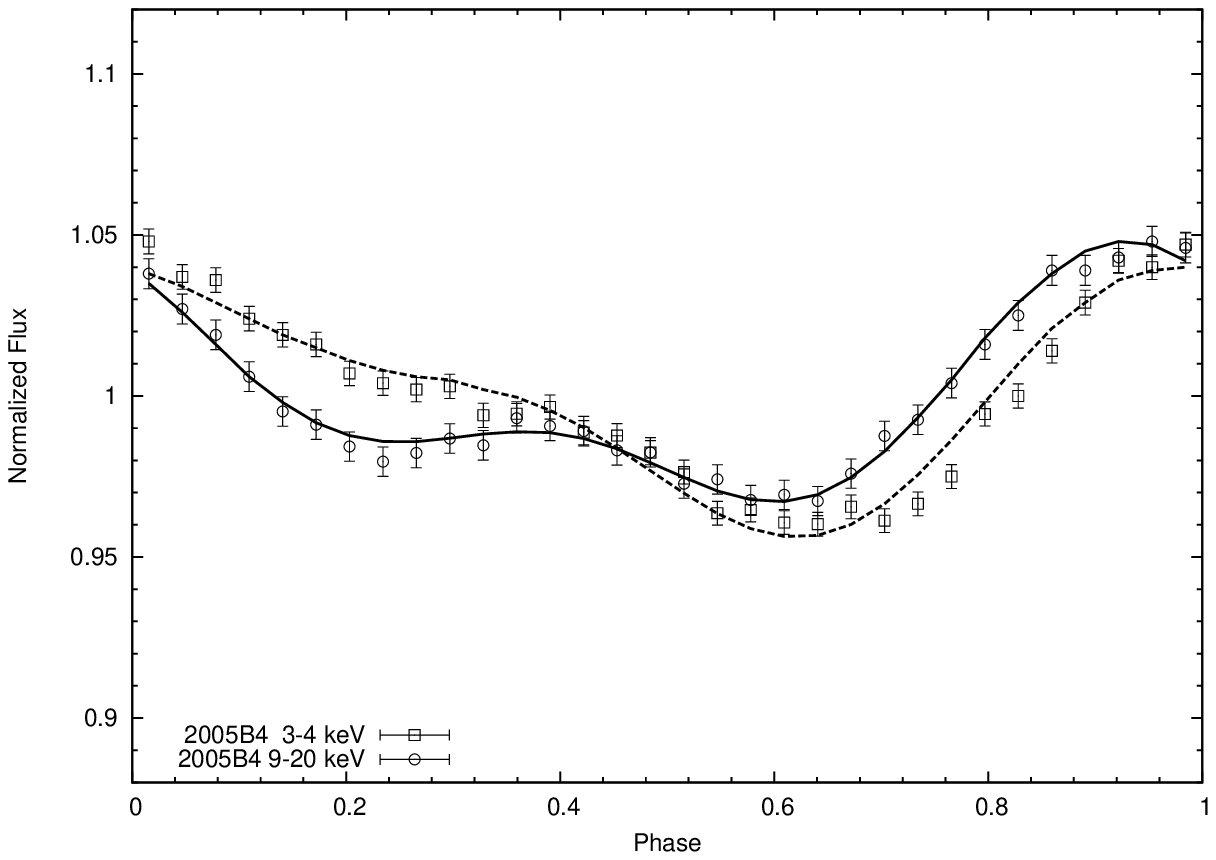}
\caption{Data from the 2005B4 period of the 2005 outburst (squares 3-4 keV, circles 9-20 keV). The solid 
and dashed curves correspond to the best-fit model that results from
a joint fit of the 1998B4 and 2005B4 data sets. This model (see row 1 of
Table \ref{tab:joint98-05B4}) has $2M/R=0.6$, $M=1.19 M_\odot$ and 
$\chi^2 = 109$ for 110 degrees of freedom.
}
\label{fig:05b4}
\end{figure}

In this paper we perform joint fits between different time periods. In order to provide the most useful
constraints on the model, we need to identify the time intervals in which the pulse shapes 
are least like each other. 
To this end, we compute reduced $\chi^2$ values comparing each of the 21 possible pairs of time intervals using both 
energy bands described above with relative normalizations a free parameter. The lowest reduced $\chi^2$ is 1.17
 (62 degrees of freedom) for the pair 2002B3 and 2005B3 showing
that the pulse shapes are not statistically different from each other. 
The next lowest reduced $\chi^2$ is 3.96 which has chance probability $<10^{-16}$. 
Thus all other pairs of pulse shapes are significantly different. For this reason, we omit the 2005B3
pulse shape from our analysis since it has larger error bars that the 2002B3 pulse shape. The 
data for the two energy bands for each of these time periods are shown in Figures~\ref{fig:98b4} -
\ref{fig:05b4}. The low-energy band data are shown using squares in Figures~\ref{fig:98b4} - \ref{fig:05b4}
while circles denote the high-energy band data. The largest 
reduced $\chi^2$ values result from comparisons of the 1998B4 data with any of the other data, ranging from
a low of 22.5 (1998B4 and 2005B1) to a high of 63.0 (1998B4 and 2005B4). The large differences can be easily
seen by eye: the pulse modulation is about 10\% for the 1998B4 data while it is only about 5\% for all
of the other time periods. Thus we find that the 1998 outburst pulse shape is significantly different from
all of the pulse shapes resulting from the 2002 and 2005 outbursts.

We now compare the data used in this paper with the data used in previous analyses
\citep{PG03,LMC08} of SAX J1808. \citet{PG03} used data from the 1998 outburst that approximately spans the 1998B4 
and 1998B5 time periods, binned into 16 time bins per period, and separated into a 3-4 keV low-energy band
and a 12-18 keV high-energy band. In the analysis by \citet{LMC08} the same data as \citet{PG03} was used, 
but a bolometric light curve constructed by \citet{Pap05} using data from a very short period (22 hours) of the 
1998B4 data was also used. 
The constraints on the mass and radius arising from the bolometric
light curve were not very strong compared to the constraints arising from the two-band light curves (see \citet{LMC08}). 
In the present analysis, we require time periods over which the pulse shape is stable, 
thus we use those  determined by \citet{Har08}.
We use pulse shapes with 32 time-bins per period because they show some features that are smoothed
out when binned down to 16 time-bins per period. The high-energy band that we use in the present analysis is
somewhat wider than the high-energy band used by \citet{LMC08}. We have confirmed that the change in the 
high-energy band does not significantly change the best-fit models of the 1998 outburst data. 
For the pulse shape analysis we added systematic errors equal to the statistical errors,
as is common practice.

\section{Hot Spot Model}
\label{s:models}

The models used in this analysis are similar to those used by  \citet{LMC08} to
model SAX J1808 and by \citet{LMCC09} to model XTE J1814-338.
(These models are similar to that of \citet{PG03} for SAX J1808, but include time-delays
and oblateness, which are important, e.g. see \cite{MLCB07}). 
The model for each pulse shape is
composed of three components: (1) Comptonized flux in the high-energy band, (2) Comptonized flux
in the low-energy band, and (3) Blackbody flux in the low energy band. For each component, the
observed flux, integrated over the energy band, is given by
\be
F_i = I_i \eta^{3+\Gamma_i} (1 - a_i \mu)
\ee
where the subscript ``i'' takes on values of 1, 2, or 3 for the 3 components, $I_i$ is a 
normalization, $\eta$ is the Doppler boost factor that depends on the pulse phase, 
$\Gamma_i$ is the effective power-law index for the component, $a_i$ is the 
anisotropy parameter, and $\mu$ is the cosine of the angle between the direction that
the photon is emitted and the normal to the surface at the location of the emitter. 
As in previous models, we take $a_3=0$ for the blackbody component, and set $a_1 = a_2 = a$ 
for the Comptonized components. The spectral model fixes the values of the 
power-law indices $\Gamma_i$ as well as the relative normalization of the low-energy
band components through the parameter $b = \bar{F_3}/\bar{F_2}$. In these models, it
is assumed that the Comptonized and Blackbody emissions both originate from the same
region on the surface of the star.

The flux in each
time-bin is computed through the oblate Schwarzschild approximation \citep{MLCB07}
which accounts for the light-bending caused by gravity, the time-delays caused by
the finite travel time for photons across the star and the oblate shape of the 
rotating star. Since the light-bending 
depends only on the 
ratio of the neutron star's mass to radius at the spot location, the
star's gravitational radius, $2GM/Rc^2$, is
kept fixed for a series of fits. (For simplicity, when referring to the 
gravitational radius, we use units with $G=c=1$.)
The lowest $\chi^2$ for a given $2M/R$ is 
found. This allows the most efficient use of ``look-up tables'' for the 
light-bending. 
Once the the ratio of $2M/R$ has been specified,
there are eight free parameters for a fit to one pair of 2-energy band 
pulse shapes (say 1998B4): 
R, $\theta$, $i$, $I_1$, $I_2$, $b$, $a$, and $\phi$.
The angle $\phi$ is an arbitrary phase.  The angle
between the spot's centre and the spin axis is $\theta$, and the observer's
inclination angle is $i$.
We assume that the orbital and spin angular momenta are aligned, so that
the inclination angle $i$ is the same as the binary's inclination angle. 
Since the angles $\theta$ and $i$ are both measured from the ``north''
spin axis, an angle smaller than $90^\circ$ corresponds to a spot or 
observer in the northern hemisphere, while an angle larger than $90^\circ$
corresponds to the southern hemisphere. We expect that it is very 
unlikely that the spot and observer could be in opposite hemispheres,
so we restrict our models to ones where $\theta$ and $i$ are  both
less than $90^\circ$.  The parameters $a$ and $b$ are not freely
varied:  the anisotropy is restricted to $0 < a < 1$ and the parameter $b$ is 
restricted to be within 1 sigma of the value given by the spectral model.

In this paper we also perform simultaneous fits to the data from two time periods. In 
these simultaneous fits, the parameters $M, R$, and $i$ are assumed to have the same
value in both time periods. All other parameters are allowed to change their values, 
subject to the restrictions on $\theta, a$ and $b$ mentioned in the previous paragraph. This 
brings the number of free parameters to 14.

\subsection{Effect of Spot Size, Shape and Number}
\label{size}

In this paper we restrict the models to only one visible
spot. 
Prior to doing this we carried out fits to the data from all seven time periods 
with a 2-spot model. The best fit 2-spot models in all cases had zero amplitude
for the second spot, indicating that no second spot is required by the observed
pulse shapes.
This is consistent with the model introduced by \citet{IP09} where the accretion
disk hides the antipodal spot on the other side of the star during the early stages of
the outburst (stages 3 and 4 in the terminology of \citet{Har08}; stages P and SD in the terminology
of \citet{IP09}). In the later periods of the burst (stages 5 and 6) the disk recedes away from
the star in the \citet{IP09} model, exposing the antipodal spot and creating a more complicated
pulse shape. The low flux and signal-to-noise in the pulse shapes corresponding to stages 5 and 
6 of the 2002 and 2005 outbursts mean that it is not feasible to use these pulse shapes in the 
present analysis.

In an earlier paper \citep{LMCC09}, the light curve of a different neutron star, 
XTE~1814, was analyzed. In that case a one-spot model led to a residual (or bump) in the 
pulse shape that was seen in all wavelengths. In that case, the residual provided the motivation for a two-spot model
for the data which did indeed fit the data significantly better. In the case of SAX J1808, the one-spot
models (shown in Section \ref{s:results}) do not lead to a significant excess feature at all wavelengths 
that would motivate a two-spot model.

\citet{PG03} have shown using analytical approximations that the theoretical pulse
shape resulting from a hot spot is independent of the spot size, as long as the
spot radius is small compared to the size of the neutron star radius. Previous 
computations of pulse shapes (e.g. \citet{LMC08};\cite{LMCC09}) have confirmed that the assumed
spot size does not strongly affect the resulting pulse shape. For this reason, we use
a simplified computational procedure that treats the spot size as infinitesimally small. 
This speeds up the computations considerably compared to the use of spots that cover
a larger area. Since the distance to the star and the overall intensity
normalization are kept arbitrary, the area of the spot is not needed in our models.
 
The best-fit models using one infinitesimal spot have unphysically low masses. For example
in row 1 of Table \ref{tab:spot} we show the best-fit model for the 2002B3 epoch with 
$2M/R$ fixed at $0.5$. (Results for other values of the gravitational radius are similar.)
Since there are 2 energy 
bands with 32 time-bins each and a total of 8 free parameters, there are
56 degrees of freedom.
The best fit model has a chi-squared value of 69.7 for 56 degrees of freedom, 
and  a mass of only 0.3 $M_\odot$ which is unreasonably low
for neutron stars, or for quark stars. Increasing
the radius of the spot so that it has a radius of $0.1 R$ or $0.2 R$ does not change 
the mass or $\chi^2$ significantly, as can be seen in rows 2 and 3 of Table \ref{tab:spot}.

We also tested the hypothesis that a non-circular spot could affect the results. We tested 
a long, thin uniform intensity spot that had a length of $0.4 R$ azimuthally. The result for this extreme 
shaped spot is shown in row 4 of Table \ref{tab:spot} where it can be seen that there is 
again no significant change from the infinitesimal spot. A similar test of a long, thin
spot with only latitudinal extent gives similar results. An alterate type of non-circular
spot can be approximated by a two spot model where both spots are artifically forced to have
the same intensity, but the relative angular locations are allowed to be free. This would 
allow for two spots to be close together, giving an effective irregular shape. The result for 
this two-spot model is shown in row 5 of Table \ref{tab:spot}. This model marginally
improves the $\chi^2$, but the best-fit mass is still only $0.354 M_\odot$. We 
conclude that changes in the spot size and shape do not help the models to 
converge to physical values for neutron stars.

\subsection{Scattered-Light Model}
\label{s:disk}

The hot-spot model described in section \ref{s:models} was used to fit two different 
data sets simultaneously. The resulting fits were found to be very poor. For 
example, data from 1998B4 and 2002B3 were fit using this model. The case 
for $2M/R = 0.5$ is shown in the first row of Table \ref{tab:sc1} (labelled ``No Scatter'').
The resulting poor $\chi^2$ of 200 for 113 degrees of freedom is typical for
fits done for other values of $2M/R$ and for other pairs of data.

The poor fits that result from the application of the hot-spot model motivates
us to introduce an extra component to the model that includes light that travels 
from the hot-spot on the star, scatters off of material 
near the star, and then arrives at the observer. The scattering material could be 
either an accretion column or an accretion disk. If the light is scattered by the disk, then there will be 
a bright pattern on the disk that will appear to rotate around the star at
the same frequency as the star, as studied by \citet{SS01}. In this section we
will outline a simple model for scattering off of an optically thin cloud of 
electrons, such as the ionized surface layer of an accretion disk.

In this simple model, we do not include the effects of gravitational
light-bending or doppler boosting for the scattered light in order to include only the most important features.
The reason why we can neglect these normally important relativistic effects is that the amplitude of the
scattered light is a very small fraction of the amplitude of the light that travels directly from the spot to 
the observer. We have tested that the inclusion of Doppler boosting to the scattered light model does not
make any significant difference to the outcome of the fits. 

Consider a spot at co-latitude $\theta$, azimuthal angle
$\phi$ on the surface of a star of radius $R$. The unit vector pointing
from the center of the star to the spot is
\be
\hat{n} = \sin\theta (\cos\phi \hat{x} + \sin\phi \hat{y}) + \cos\theta \hat{z}.
\ee
Photons emitted by the spot are scattered by material at co-latitude $\vartheta$
and azimuthal angle $\varphi$ at a distance $r$. The unit vector pointing to 
the location of the scatterer is
\be
\hat{r} = \sin\vartheta (\cos\varphi \hat{x} + \sin\varphi \hat{y}) + \cos\vartheta \hat{z}.
\ee
When light bending is neglected, the photon travels on a straight line defined by the vector 
$\vec{\ell}= r \hat{r} - R \hat{n}$ which has magnitude  
\be
\ell = r \left( \sin^2 \sigma + (\cos\sigma - R/r)^2\right)^{1/2},
\label{eq:ell}
\ee
where $\sigma$ is the angle between the vectors $\hat{r}$ and
$\hat{n}$ defined by
\be
\cos\sigma = \cos \theta \cos \vartheta + \sin \theta \sin \vartheta \cos (\phi-\varphi).
\ee
The vector pointing towards the observer is $\hat{k} = \sin i \hat{x} + \cos i \hat{z}$,
so the photon must be scattered through an angle $\rho$ in order to be detected, 
where $\rho$ is defined by
\begin{eqnarray}
\cos \rho & = & \hat{\ell} \cdot \hat{k} = \frac{1}{\ell} \left[- R(\cos\theta \cos i + \sin\theta \sin i \cos\phi) \right.  \nonumber \\
&& \left. + r(\cos\vartheta \cos i  + \sin\vartheta \sin i \cos\varphi) \right].
\label{eq:cosrho}
\end{eqnarray}

The differential cross-section for Thompson scattering is 
$
\frac{d\sigma}{d\Omega} \propto (1 + \cos^2\rho)
$
where $\rho$ is the angle that the photon is scattered through. 
If the flux of light originating from the hot spot at angle $\phi$ that 
impacts the scatterer at $\varphi$ is 
$F(\phi)$, the flux of scattered light reaching the observer is then 
$
F_{sc} \propto F(\phi)  (1 + \cos^2\rho)
$
where the constant of proportionality depends on the optical depth of 
the scattering material.

In order to model scattering of light by an optically thin cloud of electrons close to the accretion disk, the co-latitude of
the scattering material  is set to $\vartheta = \pi/2$. Photons emitted by
the spot at angle $\phi = 2\pi t/P + \phi_0$ will impact the disk over a range of azimuthal angles $\varphi$
and radii $r$. The pattern appearing on the disk will be centered around an
angle $\varphi$ equal to $\phi$. However, the photons emitted by the spot and then scattered by the
disk reach the observer at a later time than the photons that travel directly to the observer without
scattering. A simple way to treat the time lag is to define an effective phase lag $\Delta \phi$
so that $\varphi = \phi + \Delta \phi$, where 
$\Delta \phi$ includes the delays due to the light propagation time from the star to the disk as well
as the light-crossing times for the photons to cross the disk. The lowest order expression for the phase
lag (ignoring relativistic corrections) is
\be
\Delta \phi = 2 \pi( \frac{r (\cos^2 \theta + (\sin\theta - R/r)^2)^{1/2}}{c P} - \frac{r}{c P} \sin i \cos\phi )
\label{eq:deltaphi}
\ee
where $P$ is the neutron star's spin period. The time lags for the scattered light 
should be much less than the star's spin period. Since $cP = 750$ km for \saxj, physical 
solutions should have a scattering radius $r << cP$. In this model, we are approximating the
extended pattern on the disk as an effective point source. We tested more complicated models with a 
smeared out pattern on disk and found that the changes from the point source model were insignificant.

The minimum  radius of impact on the disk for a photon emitted at co-latitude $\theta$ is $r_{min} = R/\sin\theta$. 
Similarly, parts of the disk at radii $r > r_{shadow}$ can be observed at all values of $\varphi$, where the shadow radius is
$
r_{shadow} = R/\cos i.
$
As a result, the pattern on the disk is always visible to the observer 
if $\theta \le \pi/2 - i$. If the spot is at a high latitude (ie., $\theta \ll 1$) as suggested by
\citet{Lam08}, then the illuminated part of 
the disk will be at $r \gg R$ and the illuminated part of the disk will not be eclipsed by the star.

 As the spot moves around the star, the resulting pattern on 
the disk appears to move around the star at the same rate. In the frame rotating with the star, the 
pattern will not appear to move, so the flux $F(\phi)$ impacting the disk will be a constant. As a result,
the scattered flux observed will be proportional to just the differential cross-section. In addition to the 
14 parameters required for the hot spot model introduced in the previous section, the scattered light 
model requires that we add two more free parameters: $I_{sc}$ and $r$. The scattered light is added to the 
model by adding the term 
\be
F_{sc} = I_{sc} (1 + \cos^2\rho)
\ee
to each energy band. Since we are implementing a simple model for scattering from a disk ($\vartheta = \pi/2$),
the scattered flux is
\begin{eqnarray}
F_{sc} &=& I_{sc} \left( 1 + 
 \left(\frac{r}{\ell}\right)^2 
\left[   \sin i \cos(\phi+\Delta\phi)
 \right. \right.\nonumber \\ 
&& \left. \left. - \frac{R}{r}(\cos\theta \cos i + \sin \theta \sin i \cos \phi)
\right]^2\right), \label{eq:scat-flx}
\end{eqnarray}
where $\Delta\phi$ depends on the free parameters $r, i, \theta$ through equations (\ref{eq:ell}) and (\ref{eq:deltaphi}). 
The normalization for the hot spot model is defined so that the amplitudes $I_1$ and $I_2$
are of order unity, so a physical scattered light model should have an amplitude $I_{sc} << 1$. 

In row 2 of table \ref{tab:sc1} the results of a joint fit to the 1998B4 and 2002B3 data sets 
using the hot spot model along with the
simple scattering model given by equation (\ref{eq:scat-flx}) are shown (labeled ``model 1''). 
Three new parameters were introduced, the flux amplitudes $I_{s1}, I_{s2}$ for the scattered light in 1998B4 and 
2002B3 and $r$, the effective radius at which the scattering mainly occurs. The addition of these 3 parameters
to the model bring the $\chi^2$ value down to 120, which is an acceptable fit for 110 degrees of freedom. The 
resulting best-fit stellar model has a mass of $1.44 M_\odot$. The best-fit spot latitudes are close to the North pole
in both time periods, while the inclination angle is large, placing the observer close to the equatorial plane of the 
star. The best-fit scattering radius of 116 km satisfies both inequalities $r > R/\sin\theta_{1,2}$ and $r > R/\cos i$
so that this location will be illuminated by the spot and will always be visible to the observer. The relative 
amplitude of the scattered radiation in 1998B4 is $0.2\%$ of the flux coming directly from the spot. Similarly, the 
relative amplitude for the 2002B3 data is $0.8\%$. Both amplitudes are very small fractions of the direct flux, as 
would be expected for scattered light. The minimum in $\chi^2$ is fairly broad in the parameters 
corresponding the mass of the star and inclination, with the one-sigma allowed region including a range of masses
between 1.2 - 1.6 $M_\odot$.
Only one model (for $2M/R=0.5$) is shown in table \ref{tab:sc1}, but the results are similar for
other values of $2M/R$.

The best-fit solutions using the scattering model described above all have the property that the location
of the scatterer is large compared to the radius of the star, $r \gg R$. This motivates the approximation
of equation (\ref{eq:scat-flx}) to lowest order in $R/r$, 
\be
F_{sc} = I_{sc} \left( 1 + \sin^2 i \cos^2(\phi+\Delta\phi) \right).
\ee
Furthermore, the inclination angles are fairly close to 90 degrees, so $\sin i \simeq 1$. Using this approximation
leads to
\be
F_{sc} = I_{sc} \left( 1 + \cos^2(\phi+\Delta\phi) \right).
\label{eq:flux2}
\ee
If the best-fit value of $r/R \gg 1$ and $\sin i \sim 1$, equation (\ref{eq:flux2}) should be a good approximation
to equation (\ref{eq:scat-flx}).

As an alternative to the scattered light model given by equation (\ref{eq:scat-flx}), we now test the simpler
model given by equation (\ref{eq:flux2}), with $\Delta \phi$ as a free parameter, instead of $r$. 
 The best-fit stellar model that results when equation (\ref{eq:flux2}) 
is added to the basic hot-spot model and fit to the 1998B4 and 2002B3 data sets is shown in row 3 of 
table~\ref{tab:sc1} (labelled ``Model 2''). The best-fit star (with $2M/R=0.5$) using Model 2 has  $\chi^2 = 120$
for 110 degrees of freedom, so it is as good a fit as Model 1.
The best-fit values of the various parameters are different from the values for Model 1, however, the differences are
smaller than the one-sigma differences allowed by either model.
For instance the mass using Model 2 is 1.57 $M_\odot$, which lies in the one-sigma region allowed by Model 1. 
We conclude that the two models
produce results that are statistically the same.
For this reason, we use the simpler
model using
equation (\ref{eq:flux2})  to model the scattered light for the rest of this study.

Since the addition of 3 parameters ($I_{s1,2}$ and $\Delta\phi$) brings about a significant reduction
in $\chi^2$, and the masses are raised to realistic values, we include this scattered-light model 
in subsequent fitting of the pulse shapes.
In summary, this component is required in order to obtain acceptable fits and to
obtain physically reasonable masses. 

A small constant (or ``DC'') flux component was required in a similar analysis
of XTE J1814 (see section 3.2 of \citet{LMCC09}). In the case of XTE J1814, 2 free parameters (one for
each energy band) were included to improve the quality of the fits. The amplitudes
of the DC components were less than 2\% of the pulsed component, similar to what is seen in the
present analysis of SAX~J1808.

\section{Results}
\label{s:results}

A previous pulse-shape analysis
\citep{LMC08} of the 1998 outburst led to a very low best-fit mass and radius for the 
neutron star. Although the 1998B4 data are taken over a slightly different time period 
than the data used by \citet{LMC08}, a reanalysis of the 1998B4 data does not change
the best fit mass and radius by very much. For instance, Table 5 of \citet{LMC08} shows
the best fit model has a neutron star mass of $0.96 M_\odot$ and $\chi^2/\hbox{dof} = 36.3/24$.
When we fit the 1998B4 data including the scattered-light model, we find a best fit mass
of $0.98 M_\odot$ and $\chi^2/\hbox{dof} = 63.3/54$, which is not significantly different.

An alternative strategy for fitting the data, which should reduce the effects of
any near-degeneracies in model parameters, is to do a joint
fit between the 1998B4 data and the other data segments. As noted in section \ref{s:data},
the 1998B4 data has a much higher modulation than the other data segments. In addition,
the error bars for the 1998B4 data are relatively small and the data has fairly low scatter
compared to most of the other data segments. In the $\chi^2$ comparison between the 
various data streams described in section \ref{s:data} we note that the largest differences
in pulse shape can be found when comparing the 1998B4 data to any of the other data streams.
For these reasons it makes sense to attempt to jointly fit the 1998B4 data with the other
streams.

For all joint fits we consider two data streams at a time, 1998B4 and one other data set.
Due to the large number of parameters needed to model the data, it is not computationally 
feasible for us to jointly fit more than 2 data streams at a time.
In a joint fit, the mass, radius and inclination angle of the neutron star are assumed to be
identical in all years. The angular location of the spot $\theta$, anisotropy parameter $a$,
blackbody to Compton ratio $b$, scattered-light parameters, normalizations and phases are all allowed to 
vary with time. This leads to a total of 18 parameters in the joint fits. Since there are 4 
light curves (2 energy bands for each time period) with 32 points each, the total degrees
of freedom are 110.  

\subsection{Joint Fits to 1998B4 and 2002B3 Data}

The joint analysis of the 2002B3 data with the 1998B4 data results in best-fit neutron star
models with larger masses compared to the best-fit models that only use the 1998B4 data. 
Table \ref{tab:joint98-02B3} shows the values of the best-fit parameters for a range of
neutron star compactness values ($2M/R$). Each row of Table \ref{tab:joint98-02B3} is
generated by keeping the compactness ratio $2M/R$ (at the location of the spot) 
at a fixed value and varying all other
parameters until the lowest value of $\chi^2$ is found.  For each value of $2M/R$ the 
following variables are displayed: the mass of the neutron star $M$, the equatorial 
radius of the neutron star $R$, the angular location of the spot $\theta_1$ (as measured from the
North pole) in 1998, the angular location of the spot $\theta_2$ in 2002, the 
inclination angle $i$, the anisotropy parameters for 1998 ($a_1$) and 2002 ($a_2$). 
We assume the orbital and spin axes are aligned, so $i$ is also the binary's inclination 
angle.
In addition there are 4 normalizations, 2 phases, 2 blackbody to Compton ratios,
and 4 scattered-light parameters which are not displayed in the table, since they are of 
no interest here. A $\chi^2$ penalty was used to keep the blackbody to Compton
ratios $b_1$ and $b_2$ within 1 sigma of the spectral models. 

The best-fit models in Table \ref{tab:joint98-02B3}  should be compared with Table 5 of
\citet{LMC08} where only data for 1998 was used. Consider the best-fit models with $2M/R=0.4$ 
in both tables. When only 1998 data was used, the best-fit model had small mass and radius 
($M=0.90 M_\odot$ and $R=6.7$ km) and both the spot's angular location and the inclination
angle were close to 30 degrees. The addition of data from 2002B3 allows for a larger mass
and radius ($M=1.43 M_\odot$ and $R=10.8$ km) for the same value of compactness. The larger
radius is compensated for by moving the spot closer to the spin axis with $\theta_1$ close to 
10 degrees in 1998 and 5 degrees in 2002B3. 

The best-fit values of the scattered-light amplitudes $A_1$ are $2\times 10^{-3}$ for 1998B4 
and $8\times10^{-3}$ for 2002B3, where the flux is normalized to 1. 

The best-fit model with $2M/R=0.6$ in Table \ref{tab:joint98-02B3} is shown 
in Figures \ref{fig:98b4} and \ref{fig:02b3} with dashed curves for the low-energy
bands and solid curves for the high-energy bands. All four light curves are fit
simultaneously, but for clarity, the data and best-fit models are plotted in two separate 
figures. The relative phases between the high and low energy-bands for each data set are 
included in the models.


\begin{figure}
\plotone{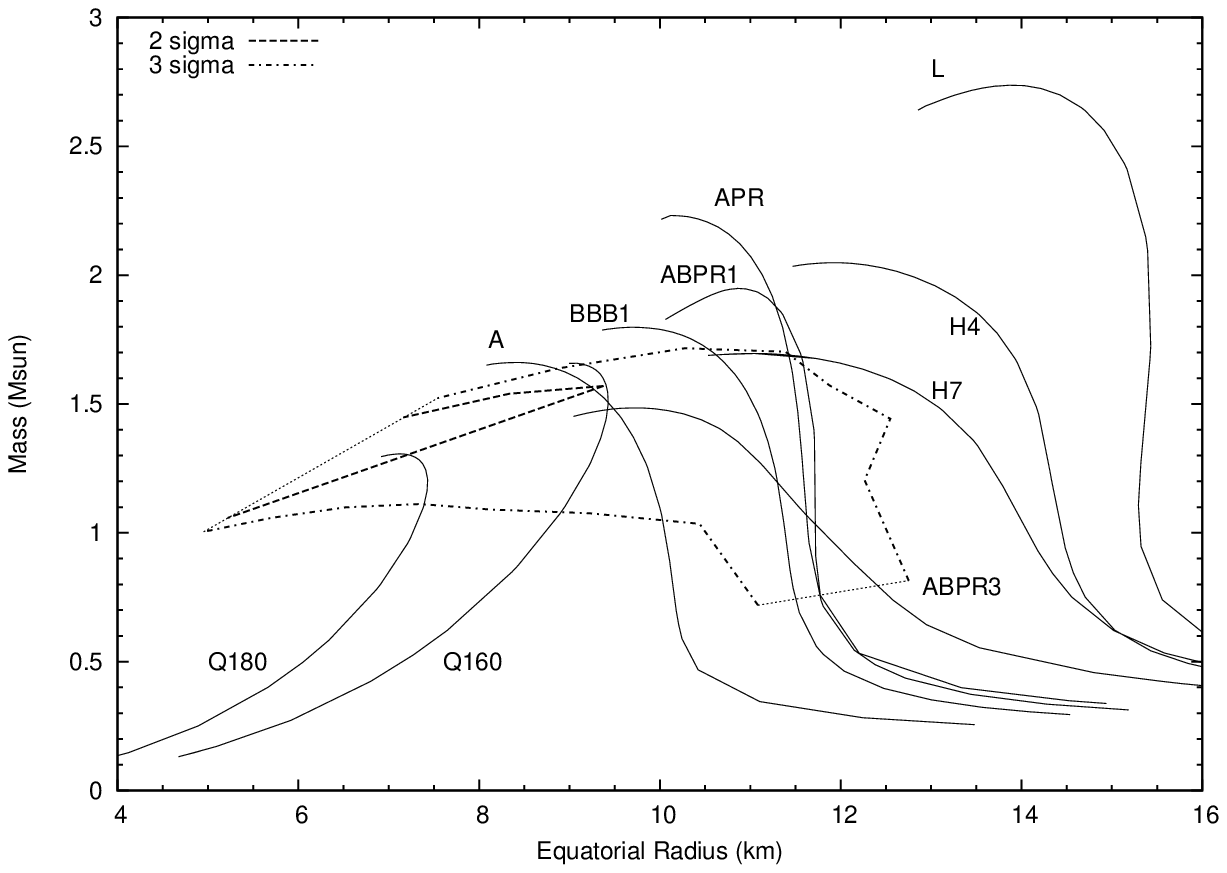}
\caption{Mass and Radius confidence contours for joint fits to 1998B4 and 2002B3 data using the
constraint $i \le 90^\circ$. Mass-radius curves (solid curves) for stars spinning at a frequency of 401 Hz are
shown for a variety of EOS. Equations of state displayed are: \citet{AB77} models A
\citep{eosA} and L \citep{eosL}; mixed phase quark-hadron stars ABPR1-3 \citep{ABPR}; 
APR \citep{APR}; BBB1 \citep{BBB}; hyperon stars H4 and H7 \citep{LNO};
and quark stars Q160 and Q180 (where the number corresponds to value of $B^{1/4}$ in MeV
where $B$ is the bag constant \cite{Gle00}). 
Confidence contours for $2\sigma$ (dashed curve) and $3\sigma$ (dot-dashed curve) are shown. 
The boundaries shown as dotted lines correspond to the largest (0.6) and smallest (0.2) values 
of $2M/R$ used in the computations.
}
\label{fig:mr90}
\end{figure}

Figure \ref{fig:mr90} shows 2 and 3$\sigma$ confidence contours on the mass-radius plane for 
joint fits with 1998B4 and 2002B3. This figure is generated by specifying a mass and then 
varying all parameters until the lowest value of $\chi^2$ for the mass is found. Two constraints
were added to the variation of parameters. First, the ratio $2M/R$ was only sampled in the range 
$0.2 \le 2M/R \le 0.6$ in order to stay within the bounds of reasonable equations of state. 
Second, an unconstrained variation of the inclination angle leads to acceptable fits for solutions 
with $i > 90^\circ$, which is very unlikely. Hence, we constrained the inclination angle to values 
with $i \le 90^\circ$. The resulting confidence contours in Figure \ref{fig:mr90} should be 
compared with the corresponding confidence contours for the 1998 data shown with bold curves in Figure 3 of \citet{LMC08} 
(labelled ``oblate and time delays''). The 3$\sigma$ confidence region only allows very small stars 
with $R < 7$ km if only the 1998 epoch is used. The additional data from 2002B3 and allowance for scattered-light used in the present
analysis allows stars with larger radii, extending to 13 km at the 3$\sigma$ confidence level. 
The largest mass stars allowed (at 3$\sigma$) have $M = 1.7 M_\odot$. 
At 3$\sigma$, the data allow for a wide variety of modern equations of state.

\subsection{Inclination Constraints}
\label{s:incl}

\begin{figure}
\plotone{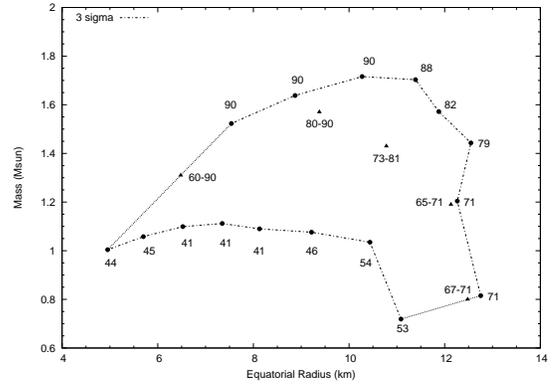}
\caption{Inclination ranges for the joint fit to 1998B4 and 2002B3 data sets. 
Points labeled with a triangle correspond to the best-fit models shown in 
Table~\ref{tab:joint98-02B3}. For each one of these points, the
$3\sigma$  range of 
inclination angles (in degrees) for the given mass and radius is shown.
The dot-dashed curve shows the same 3$\sigma$ limit on the allowed values of 
mass and radius shown in Figure~\ref{fig:mr90}. Selected points along 
the $3\sigma$ contour are shown with circles. The inclination 
angle corresponding for each of these points is shown.
}
\label{fig:inc}
\end{figure}

The best-fit models displayed in Table~\ref{tab:joint98-02B3} have inclinations ranging
from 70 - 90 degrees. However, for each best-fit set of values for mass and radius shown
in Table~\ref{tab:joint98-02B3}, there is a range of inclination angles allowed by the data.
The range of inclination angles allowed at the 3 $\sigma$ level is found by keeping the
mass and radius fixed at the values shown in the table, and allowing all other parameters
to vary. In Figure~\ref{fig:inc}, the locations on the mass-radius plane for the five models
shown in Table~\ref{tab:joint98-02B3} are marked with triangles. The allowed 
$3 \sigma$ range of inclination angles (in degrees)
 for each of these points is shown on Figure~\ref{fig:inc}. For example, the
model with $M =1.43 M_\odot$ and $R=  10.8$ km is allowed
inclination angles in the range of $73^\circ < i < 81^\circ$, although the lowest $\chi^2$
results for $i=75.5^\circ$.

On Figure~\ref{fig:inc} the same 3 $\sigma$ contours on the mass-radius plane that were
shown in Figure~\ref{fig:mr90} are redisplayed. For selected points on the contours (shown
with circles), the one value of inclination  that is allowed is shown.  For example, 
Figure~\ref{fig:inc} shows that the data allows at the 3 $\sigma$ confidence level 
a star with $M=1.44 M_\odot$,  $R = 12.5$ km and $i=79^\circ$.  In general, the lowest mass
stars require smaller inclination angles (near 40 degrees) while the highest mass stars 
require inclinations that are as high as is allowed.


There are a number of observations that suggest that the inclination angle may be smaller than 90 degrees.
The strongest evidence is the absence of eclipses, which constrains
$i<82^\circ$ assuming a Roche lobe filling companion \citep{CM98}.
\citet{Wang}
observed V4580 Sgr, the optical counterpart of SAX J1808, during the 1998 outburst. They modelled the
optical and IR data using an X-ray-heated accretion disk and found limits on the inclination of
$ 42^\circ < i < 88^\circ$ if the distance is $d = 2$ kpc and $20^\circ < i < 65^\circ$ if $d = 3$ kpc.
Further modelling of the optical data observed during quiessence by \citet{Wang09} suggests that
$i < 70^\circ$ using the distance of $d = 3.5$ kpc derived by \citet{GC06}. \citet{Deloye08} 
modelled the optical data from quiessence using a model that includes emission from both
the disk and the companion. Using a distance of 3.5 kpc, \citet{Deloye08} constrain the
inclination to $32^\circ < i < 74^\circ$ if 10\% uncertainty in the distance is assumed. 

A feature interpreted as an iron line relativistically broadened through the orbital
motion of the disk was detected in X-ray observations during the 2008 outburst 
by \citet{Cac09} and \citet{Pap09}. Both groups modelled the emission feature
and constrained the inclination angle. \citet{Cac09} found $i = 55^{+8}_{-4}$ degrees,
while \citet{Pap09} provide a less precise constraint of approximately $i > 60^\circ$.
A different type of constraint on the inclination arises from the analysis of the
2002 outburst data by \citet{IP09}. In their analysis, \citet{IP09} assume that 
the X-ray emission is produced by two antipodal spots, one of which is sometimes
hidden by the accretion disk, due to movement of the inner edge of the accretion disk.
In this model, they find that the data can be explained by inclinations in the
range $i < 73^\circ$. 

Although all of the constraints on inclination have some model dependence, there is a 
general agreement that inclinations smaller than about $70^\circ$ are suggested by the 
different data. With this in mind, we refit the data for 1998B4 and 2002B3 subject to
the constraint $i\le70^\circ$. The best fit models for fixed values of $2M/R$ are shown in
Table \ref{tab:i70}. Comparing Table \ref{tab:i70} with Table \ref{tab:joint98-02B3}
it can be seen the the most significant change is that lower mass stars are allowed when the
inclination angle is constrained to $i\le70^\circ$. The values of $\chi^2$ increase, but
the magnitude of the increase is not significant.


\begin{figure}
\plotone{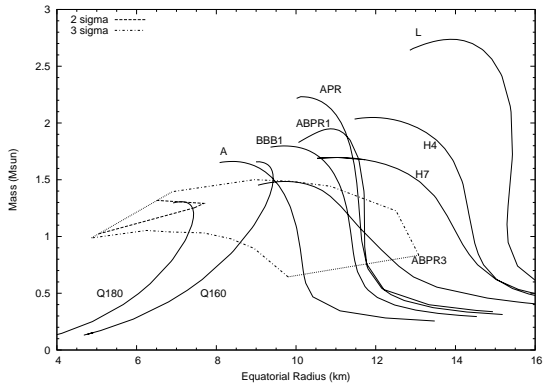}
\caption{Mass and Radius confidence contours for joint fits to 1998B4 and 2002B3 data using the
constraint $i \le 70^\circ$. Mass-radius curves (solid curves) for stars spinning at a frequency of 401 Hz are
displayed with solid curves. The radius corresponds to the equatorial radius of the star.
EOS labels are the same as in Figure \ref{fig:mr90}.
Confidence contours for 2 $\sigma$ (dashed curve) and 3 $\sigma$ (dot-dashed curve) are shown. 
}
\label{fig:mr70}
\end{figure}

Figure \ref{fig:mr70} shows the 2 and 3-$\sigma$ confidence contour on the mass-radius plane using the
constraint $i\le70^\circ$ for joint fits to the 1998B4 and 2002B3 data. The 3 $\sigma$ confidence region
shrinks when the constraint on inclination is added, however stars as massive as $1.5 M_\odot$ are still
allowed. 

\subsection{Location of the Hot Spot}
\label{s:location}

In all of the hot spot models for the 1998B4 and 2002B3 data sets, the colatitude of the spot $\theta$ is
larger (further from the pole) in 1998 compared to the colatitude in 2002. This trend is seen in all models
fitting the other data sets as well. Since a hot spot located close to the spin axis produces a less-modulated
pulse shape, this is a natural outcome of the fact that the 1998 data is highly modulated ($\sim 10\%$) and
the data for the later outbursts have smaller modulation ($\sim 5\%$). 

In the best-fit models for the 1998B4 and 2002B3 data shown in Table \ref{tab:joint98-02B3}
the shift in colatitude $\Delta \theta = \theta_1 - \theta_2$ ranges from $5^\circ$ to $8^\circ$.
When all models that fit the data within 3 $\sigma$ confidence are considered, the colatitude 
in 1998 lies in the range $9^\circ < \theta_1 < 32^\circ$, the colatitude in 2002B3
lies in the range $4^\circ < \theta_2 < 17^\circ$. The 3 $\sigma$ range of colatitude 
difference is in the range $6^\circ < \Delta \theta < 15^\circ$.  This
wandering of the hot spot location is consistent with the model proposed by \citet{Lam08} 
where a hot spot close to the spin axis moves due to movement of the star's magnetic field.

\subsection{Joint Fits to 1998B4 and 2002B4 Data}

The 2002B4 and 1998B4 data have similar pulse shapes 
(although a statistical test shows that they are different at high significance - chance probability 
$<10^{-20}$). 
Of all the data sets, these two epochs have the smallest errorbars which are roughly equal in
magnitude.
The most important difference is in
the modulation of the pulse shapes.

The best-fit models for joint fits to the 1998B4 and 2002B4 data sets are shown in 
Table~\ref{tab:joint98-02B4}. In this table, the subscript ``2'' corresponds to values
of in the 2002B4 epoch, while the subscript ``1'' corresponds to the 1998B4 epoch.
These best-fit models should be compared with the best-fit
models shown in Table~\ref{tab:joint98-02B3}. The range of masses and radii found when the
2002B4 data are fit is similar to those found when data from 2002B3 are used. The values
of mass and radius shown in Table~\ref{tab:joint98-02B4} are consistent with the 
$3\sigma$ limits arising from the joint fit between the 1998B4 and 2002B3 data, with the
exception of the model with a mass of $0.66 M_\odot$. 
 The best-fit values of the 1998 angular location
of the spot $\theta_1$ and the inclination angle $i$ differ in the best-fit models 
displayed in Tables~\ref{tab:joint98-02B3} and \ref{tab:joint98-02B4}. This is not 
surprising, since as we have shown earlier, given values of mass and radius, a wide range
of angles $\theta$ and $i$ will give acceptable fits to the data.

The best-fit model's light curve for the 2002B4 data is shown in Figure~\ref{fig:02b4}.
(The light curves for 1998B4 generated by the best-fit model are not 
distinguishable from the curves shown in Figure~\ref{fig:98b4} and are not plotted.)

\subsection{2005 Outburst Data}

The 2005 outburst was only about 70\% as luminous as the 2002 outburst \citep{Har08}. 
In addition, the RXTE satellite has lost
sensitivity since 2002. As a result, the 2005 data is much noisier and has larger error bars than the 1998 and 
2002 data as can be seen in Figures~\ref{fig:05b1}-\ref{fig:05b4}. One-spot models were used to simultaneously fit 
the 2005 data with the 1998 data. The best-fit model parameters are shown in 
Tables~\ref{tab:joint98-05B1}-\ref{tab:joint98-05B4} and the best-fit light curves are shown in
Figures~\ref{fig:05b1}-\ref{fig:05b4}.

The best-fit models that make use of the 2005 outburst data shown in Tables~\ref{tab:joint98-05B1}
-~\ref{tab:joint98-05B4} are very similar to the best-fit models computed using the 2002B4 data
and displayed in Table~\ref{tab:joint98-02B4}. 
The models with $2M/R = 0.4, 0.5$ and 0.6 (and with $M > 1.0 M_\odot$)
are consistent with the $3\sigma$ limits arising from the joint 1998B4-2002B3 fits.

The 2008 outburst was almost as bright as the 2005 outburst, however the 
degradation of the RXTE PCA over the intervening 3 years has the result
that the pulse shapes from the 2008 outburst \citep{Har09}, while similar
to the 2005 pulse shapes, are even noisier than the 2005 pulse profiles.
For this reason, we have not attempted to fit the 2008 pulse profiles.

\section{Discussion}

In this paper we analyze multiple pulse shapes resulting
from the different epochs of the 1998, 2002 and 2005 outbursts of SAX J1808. We 
jointly fit data from 1998 and several other epochs in order to find 
consistent models for the neutron star's mass, radius and inclination. 

In each joint fit to the data from two different epochs, the mass,
equatorial radius and the inclination angle are kept fixed between epochs,
while the spot location, anisotropy and blackbody-to-compton ratios
are allowed to change from epoch to epoch. We use a spectral model, motivated
by \citet{GDB02} and \citet{PG03}, that includes
blackbody radiation and Comptonized radiation in the low-energy (3-4 keV) band
and only Comptonized radiation in the high-energy (9-20 keV) band. 
We find that an extra scattered-light component is required to fit all of the pulse shapes.

The inclusion of the scattered radiation model (Section \ref{s:disk}), 
together with a hot spot that
wanders a little in its co-latitude, allows a consistent fit for all
different epochs with a consistent mass. Increased radius or increased
spot co-latitude increases the second harmonic content in the pulse shape;
scattered radiation reduces it; and increased light bending (smaller star
at a given mass) can increase or decrease second harmonic content depending
on viewing geometry. Without including all these physical effects,
the fits to some epochs require a small star  whereas other epochs require
a larger star. As we have shown, by including all
effects, we can consistently fit different epochs with a consistent mass.

The joint fit of data from the 1998B4 and 2002B3 epochs using the constraint $i<90^\circ$
leads to the $3\sigma$ confidence constraint on the neutron star mass $0.81 M_\odot < M < 1.72 M_\odot$
and equatorial radius $5.0 \km < R < 12.7 \km$, as summarized in Figure~\ref{fig:mr90}.  Inclinations as low 
as $41^\circ$ are allowed.  If the inclination is further constrained to $i<70^\circ$, as suggested by many separate 
observations, the $3\sigma$ limits are $ 0.83 M_\odot < M < 1.50 M_\odot$ and
$ 7 \km < R < 13 \km$, as summarized in Figure~\ref{fig:mr70}. Inclinations as low as $35^\circ$ are allowed in this case.

The joint fits between 1998B4 and the other epochs (2002B4, 2005B1, 2005B2 and 2005B4) yield 
independent best-fit models summarized in Tables~\ref{tab:joint98-02B4}-\ref{tab:joint98-05B4}. 
The best-fit models from these epochs with $M > 1.0 M_\odot$ have mass and radius values that
lie within the $3\sigma$ confidence region for the 1998B4-2002B3 joint fits. This indicates
that the various joint fits between the different epochs give consistent results. 

The $3\sigma$ confidence region for the 1998B4-2002B3 joint fits allows a much wider range
of masses and radii for the neutron star than a similar fit that only made use of data from 
the 1998 outburst \citep{LMC08}. In the previous analysis by \citet{LMC08} a similar type of
analysis using 2 narrow energy bands constrained the neutron star mass and radius to very small
values with $R<7$ km. The inclusion of a scattered-light component and use of joint fits with 
data from other epochs (2002 and 2005 outbursts) results in stars with
larger masses and radii than are given by fits to the 1998
data alone without the scattered-light component. 

In their analysis of the 2002 data, \citet{IP09} do not attempt to find a best-fit mass and radius for SAX J1808. For some of their models they make use of a fixed mass of 1.4 $M_\odot$ and a radius of either 10 or 12 km. However, the results of their modelling (such as the magnetic moment, visibility of an antipodal spot) do not strongly depend on the assumed values of mass and radius. A mass of 1.4 $M_\odot$ and a radius in the 10-12 km range is consistent with our final results. 

Observations of SAX J1808 in quiescence by \citet{Heinke07,Heinke09} have set limits on
the luminosity. The low luminosity during quiescence implies that rapid
neutrino cooling takes place in the inner core, such as the direct URCA process. 
\citet{YP04} show that direct URCA can take place if the core density is larger
than approximately $10^{15}$ g/cm$^3$. For a standard hadronic EOS such as
APR \citep{APR}, stars with masses larger than about $1.5 M_\odot$ 
(and spinning at 400 Hz) have central
densities larger than this critical density. Softer EOS will have 
lower mass stars for the same central density. This suggests that many of the 
models allowed by the $3\sigma$ confidence region shown in Figure \ref{fig:mr90}
could have cores dense enough to allow cooling through the direct URCA process.

A pulse-shape analysis of the accreting ms-period X-ray pulsar XTE 1814-338 \citep{LMCC09}
using the similar methods as used in this paper implied that the mass and radius 
of XTE 1814-338 must be quite large. Although the $3\sigma$ regions for 
XTE 1814-338 and SAX J1808 do not overlap, the results are not necessarily 
inconsistent. It is only necessary for an EOS mass-radius curve to pass
through the regions for both stars. For example, an EOS mass-radius curve
that is slightly to the right of APR in Figure~\ref{fig:mr90} would still be allowed by
the SAX J1808 data, and would be stiff enough to be allowed by the XTE 1814 data, and also allowed by the constraints on
the slow X-ray pulsar Hercules X-1 \citep{Leahy04}.

\acknowledgments
This research was supported by grants from NSERC. 
We thank Jake Hartman for supplying us with the pulse
profiles used in this analysis, and for helpful comments
on the paper.


\clearpage


\begin{deluxetable}{lrrrrrrr}
\tablecaption{Effect of Spot Size and Shape for stellar models with $2M/R=0.50$ fit with data from 2002B3. 
\label{tab:spot}}
\tabletypesize{\scriptsize}
\tablehead{
\colhead{Model}&\colhead{$M$}&\colhead{$\chi^2/$dof}&\colhead{$\theta$}&
\colhead{$i$}&\colhead{$a$}&\colhead{$\theta_2$}&\colhead{$d\phi$}\\
\colhead{}&\colhead{$M_\odot$}&\colhead{}&\colhead{deg.}&
\colhead{deg.}&\colhead{}&\colhead{}&\colhead{}
}
\startdata
infinitesimal spot    & 0.304 &  69.7/56 &  27.0 & 36.0 &   0.50 &   &      \\ 
circle radius $0.1 R$ & 0.305 &  69.7/56 &  27.0 &  36.2 &   0.51 & & \\
circle radius $0.2 R$ & 0.307 &  69.7/56 &  27.2 &  36.6 &   0.51 && \\
azimuthal line $0.4 R$& 0.310 &  69.2/56 &  26.2 &  35.8 &   0.51 && \\
two spots (equal amp.)& 0.354 &  62.1/54 &  12.0 &  26.0 &   0.54 & 54.4 & 60.1\\
\enddata
\end{deluxetable}

\begin{deluxetable}{lrrrrrrrrrrrr}
\tablecaption{Best Fit Parameters for Joint Fits using Data from 1998B4 and 2002B3 
with and without the scattered light component for $2M/R=0.5$.\label{tab:sc1}}
\tablewidth{0pt}
\tablehead{
\colhead{} & \colhead{$M$}&\colhead{$R$}&\colhead{$\theta_1$}&
\colhead{$\theta_2$}& \colhead{$i$}&\colhead{$a_1$}&\colhead{$a_2$}&
\colhead{$I_{s1}$}&\colhead{$I_{s2}$}&\colhead{$r$}&\colhead{$\chi^2/$dof}\\
\colhead{} & \colhead{$M_\odot$}&\colhead{km}&\colhead{deg.}&
\colhead{deg.}&\colhead{deg.}&\colhead{}&\colhead{}&\colhead{}&\colhead{}&\colhead{km}&\colhead{}
}
\startdata
No Scatter & 1.02 &  6.0 &  22.6 & 10.6 &  49.5 &  0.57 &  0.49 & 0 & 0 & & 200/113\\
Model 1 & 1.44 &  8.5 &  12.0 & 5.6 & 80.0 &    0.80 &  0.68 & 0.002 & 0.008 & 116 & 120/110\\
Model 2 & 1.57 &  9.4 &  10.5 &  4.8 &  88.9 &  0.92 &  0.77 & 0.003 & 0.007& & 120/110 \\ 
\enddata
\tablecomments{Subscript ``1'' corresponds to 1998B4 values and ``2'' to
2002B3.}
\end{deluxetable}

\begin{deluxetable}{rrrrrrrrr}
\tablecaption{Best Fit Parameters for Joint Fits using Data from 1998B4 and 2002B3.\label{tab:joint98-02B3}}
\tablewidth{0pt}
\tablehead{
\colhead{$2M/R$}&\colhead{$M$}&\colhead{$R$}&\colhead{$\theta_1$}&
\colhead{$\theta_2$}&
\colhead{$i$}&\colhead{$a_1$}&\colhead{$a_2$}&\colhead{$\chi^2/$dof}\\
\colhead{}&\colhead{$M_\odot$}&\colhead{km}&\colhead{deg.}&
\colhead{deg.}&\colhead{deg.}&\colhead{}&\colhead{}&\colhead{}
}
\startdata
0.60 &  1.31 &  6.5 &  14.3 &  6.2 &  $89.6$ &  0.77 &  0.62 &  116/110 \\ 
0.50 &  1.57 &  9.4 &  10.5 &  4.8 &  $88.9$ &  0.92 &  0.77 &  120/110 \\ 
0.40 &  1.43 &  10.8 &  10.4 &  4.9 &  $75.5$ &  0.85 &  0.73 &  123/110 \\ 
0.30 &  1.19 &  12.1 &  10.0 &  4.7 &  $70.8$ &  0.89 &  0.75 &  124/110 \\
0.20 &  0.80 &  12.5 &  10.2 &  5.0 &  $70.3$ &  1.0 &  0.88 &  123/110 \\ 
\enddata
\tablecomments{Subscript ``1'' corresponds to 1998B4 values and ``2'' to
2002B3.}
\end{deluxetable}

\begin{deluxetable}{rrrrrrrrr}
\tablecaption{Best Fit Parameters for Joint Fits using Data from 1998B4 and 2002B3 with $i\le70^\circ$.\label{tab:i70}}
\tablewidth{0pt}
\tablehead{
\colhead{$2M/R$}&\colhead{$M$}&\colhead{$R$}&\colhead{$\theta_1$}&
\colhead{$\theta_2$}&
\colhead{$i$}&\colhead{$a_1$}&\colhead{$a_2$}&\colhead{$\chi^2/$dof}\\
\colhead{}&\colhead{$M_\odot$}&\colhead{km}&\colhead{deg.}&
\colhead{deg.}&\colhead{deg.}&\colhead{}&\colhead{}&\colhead{}
}
\startdata
0.60 &  1.21 &  6.0 &  18.4 &  8.3 &  70.0 &  0.64 &  0.52 &  118/110 \\ 
0.50 &  1.27 &  7.6 &  16.0 &  7.6 &  63.3 &  0.65 &  0.56 &  122/110 \\
0.40 &  1.30 &  9.7 &  12.7 &  6.1 &  64.8 &  0.72 &  0.62 &  124/110 \\ 
0.30 &  1.09 &  11.0 &  11.9 &  5.8 &  61.9 &  0.75 &  0.65 &  125/110 \\ 
0.20 &  0.80 &  12.5 &  10.2 &  5.0 &  70.0 &  0.99 &  0.87 &  123/110 \\ 
\enddata
\tablecomments{Subscript ``1'' corresponds to 1998B4 values and ``2'' to
2002B3.}
\end{deluxetable}

\begin{deluxetable}{rrrrrrrrr}
\tablecaption{Best Fit Parameters for Joint Fits using Data from 1998B4 and 2002B4.\label{tab:joint98-02B4}}
\tablewidth{0pt}
\tablehead{
\colhead{$2M/R$}&\colhead{$M$}&\colhead{$R$}&\colhead{$\theta_1$}&
\colhead{$\theta_2$}&
\colhead{$i$}&\colhead{$a_1$}&\colhead{$a_2$}&\colhead{$\chi^2/$dof}\\
\colhead{}&\colhead{$M_\odot$}&\colhead{km}&\colhead{deg.}&
\colhead{deg.}&\colhead{deg.}&\colhead{}&\colhead{}&\colhead{}
}
\startdata
0.60 &  1.27 &  6.3 &  22.3 &  18.6 &  53.1 &  0.57 &  0.65 &  105/110 \\ 
0.50 &  1.27 &  7.5 &  21.0 &  17.5 &  46.9 &  0.57 &  0.64 &  106/110 \\ 
0.40 &  1.19 &  8.9 &  20.2 &  16.6 &  41.9 &  0.57 &  0.64 &  111/110 \\ 
0.30 &  0.96 &  9.7 &  19.5 &  16.3 &  39.4 &  0.57 &  0.64 &  108/110 \\ 
0.20 &  0.66 &  10.1 &  19.2 &  16.4 &  38.3 &  0.58 &  0.65 &  109/110 \\ 
\enddata
\tablecomments{Subscript ``1'' corresponds to 1998B4 values and ``2'' to
2002B4.}
\end{deluxetable}

\begin{deluxetable}{rrrrrrrrr}
\tablecaption{Best Fit Parameters for Joint Fits using Data from 1998B4 and 2005B1.\label{tab:joint98-05B1}}
\tablewidth{0pt}
\tablehead{
\colhead{$2M/R$}&\colhead{$M$}&\colhead{$R$}&\colhead{$\theta_1$}&
\colhead{$\theta_2$}&
\colhead{$i$}&\colhead{$a_1$}&\colhead{$a_2$}&\colhead{$\chi^2/$dof}\\
\colhead{}&\colhead{$M_\odot$}&\colhead{km}&\colhead{deg.}&
\colhead{deg.}&\colhead{deg.}&\colhead{}&\colhead{}&\colhead{}
}
\startdata
0.60 &  1.36 &  6.7 &  19.6 &  19.2 &  57.2 &  0.58 &  0.66 &  111/110 \\ 
0.50 &  1.33 &  7.9 &  18.8 &  18.6 &  49.9 &  0.58 &  0.65 &  114/110 \\ 
0.40 &  1.20 &  9.0 &  18.3 &  18.1 &  44.8 &  0.58 &  0.65 &  116/110 \\ 
0.30 &  0.98 &  9.9 &  18.0 &  17.9 &  41.6 &  0.58 &  0.65 &  117/110 \\ 
0.20 &  0.67 &  10.3 &  17.9 &  17.9 &  40.1 &  0.58 &  0.66 &  119/110 \\ 
\enddata
\tablecomments{Subscript ``1'' corresponds to 1998B4 values and ``2'' to
2005B1.}
\end{deluxetable}

\begin{deluxetable}{rrrrrrrrr}
\tablecaption{Best Fit Parameters for Joint Fits using Data from 1998B4 and 2005B2.
\label{tab:joint98-05B2}}
\tablewidth{0pt}
\tablehead{
\colhead{$2M/R$}&\colhead{$M$}&\colhead{$R$}&\colhead{$\theta_1$}&
\colhead{$\theta_2$}&
\colhead{$i$}&\colhead{$a_1$}&\colhead{$a_2$}&\colhead{$\chi^2/$dof}\\
\colhead{}&\colhead{$M_\odot$}&\colhead{km}&\colhead{deg.}&
\colhead{deg.}&\colhead{deg.}&\colhead{}&\colhead{}&\colhead{}
}
\startdata
0.60 &  1.20 &  5.9 &  23.1 &  13.9 &  53.2 &  0.57 &  0.64 &  109/110 \\ 
0.50 &  1.20 &  7.1 &  21.9 &  13.1 &  46.6 &  0.57 &  0.63 &  109/110 \\ 
0.40 &  1.09 &  8.1 &  21.1 &  12.5 &  42.1 &  0.57 &  0.62 &  111/110 \\ 
0.30 &  0.90 &  9.1 &  20.5 &  12.3 &  39.0 &  0.57 &  0.62 &  113/110 \\ 
0.20 &  0.64 &  9.7 &  20.1 &  12.0 &  37.5 &  0.58 &  0.63 &  114/110 \\ 
\enddata
\tablecomments{Subscript ``1'' corresponds to 1998B4 values and ``2'' to
2005B2.}
\end{deluxetable}

\begin{deluxetable}{rrrrrrrrr}
\tablecaption{Best Fit Parameters for Joint Fits using Data from 1998B4 and 2005B4.\label{tab:joint98-05B4}}
\tablewidth{0pt}
\tablehead{
\colhead{$2M/R$}&\colhead{$M$}&\colhead{$R$}&\colhead{$\theta_1$}&
\colhead{$\theta_2$}&
\colhead{$i$}&\colhead{$a_1$}&\colhead{$a_2$}&\colhead{$\chi^2/$dof}\\
\colhead{}&\colhead{$M_\odot$}&\colhead{km}&\colhead{deg.}&
\colhead{deg.}&\colhead{deg.}&\colhead{}&\colhead{}&\colhead{}
}
\startdata
0.60 &  1.19 &  5.9 &  23.0 &  11.3 &  54.4 &  0.57 &  0.64 &  109/110 \\ 
0.50 &  1.18 &  7.0 &  21.9 &  10.7 &  47.3 &  0.57 &  0.63 &  111/110 \\ 
0.40 &  1.07 &  8.0 &  21.1 &  10.3 &  42.5 &  0.57 &  0.63 &  112/110 \\ 
0.30 &  0.89 &  8.9 &  20.3 &  9.9 &  40.0 &  0.58 &  0.63 &  114/110 \\ 
0.20 &  0.63 &  9.6 &  19.9 &  9.7 &  38.3 &  0.58 &  0.64 &  115/110 \\ 
\enddata
\tablecomments{Subscript ``1'' corresponds to 1998B4 values and ``2'' to
2005B4.}
\end{deluxetable}


\begin{thebibliography}{}
\bibitem[Akmal et al.(1998)]{APR}
Akmal, A., Pandharipande, V. R., \& Ravenhall, D. G. 1998,
Phys. Rev. C, 58, 1804
\bibitem[Alford et al.(2005)]{ABPR}
Alford, M., Braby, M., Paris, M., \& Reddy, S. 2005,
ApJ 629, 969
\bibitem[Arnett \& Bowers(1977)]{AB77}
Arnett, W.~D., \& Bowers, R.~L. 1977, \apjs, 33, 415
\bibitem[Baldo et al.(1997)]{BBB}
Baldo, M., Bombaci, I., \& Burgio, G. F., 1997, \aap, 328, 274
\bibitem[Cackett et al.(2009)]{Cac09}
Cackett, E.~M., 
Altamirano, D., Patruno, A., Miller, J.~M., Reynolds, M., Linares, M., 
\& Wijnands, R.\ 2009, \apjl, 694, L21 
\bibitem[Cadeau et al.(2005)]{CLM05}
Cadeau, C., Leahy, D.~A., \& Morsink, S.~M. 2005, \apj, 618, 451
\bibitem[Cadeau et al.(2007)]{CMLC07}
Cadeau, C., Morsink, S. M., Leahy, D. A., \& Campbell, S. S. 2007,
\apj, 654, 458
%
\bibitem[Chakrabarty \& Morgan(1998)]{CM98}
Chakrabarty, D. \& Morgan, E.~H. 1998, \nat, 394, 346
%
\bibitem[Deloye et al.(2008)]{Deloye08} Deloye, C.~J., Heinke, 
C.~O., Taam, R.~E., \& Jonker, P.~G.\ 2008, \mnras, 391, 1619 
\bibitem[Galloway \& Cumming(2006)]{GC06} Galloway, D.~K., 
\& Cumming, A.\ 2006, \apj, 652, 559 
\bibitem[Gierli\'{n}ski et al.(2002)]{GDB02}
Gierli\'{n}ski, M., Done, C., \& Barret, D. 2002, 
\mnras, 331, 141
\bibitem[Glendenning(2000)]{Gle00}
        Glendenning, G. K. 2000, {\em Compact Stars},
        (Springer-Verlag, New York)
\bibitem[Hartman et al.(2008)]{Har08}
Hartman, J. M., et al. 2008, \apj, 675, 1468
\bibitem[Hartman et al.(2009)]{Har09}
Hartman, J. M., et al. 2009, \apj, 
702, 1673
\bibitem[Heinke et al.(2007)]{Heinke07} Heinke, C.~O., Jonker, 
P.~G., Wijnands, R., \& Taam, R.~E.\ 2007, \apj, 660, 1424 
\bibitem[Heinke et al.(2009)]{Heinke09} Heinke, C.~O., Jonker, 
P.~G., Wijnands, R., Deloye, C.~J., \& Taam, R.~E.\ 2009, \apj, 691, 1035 
\bibitem[Ibragimov \& Poutanen(2009)]{IP09}
Ibragimov, A., \& Poutanen, J. 2009, \mnras, 
400, 492
\bibitem[Lackey, Nayyar, \& Owen(2006)]{LNO}
Lackey, B.~D., Nayyar, M., \& Owen, B.~J.\ 2006, \prd, 73, 024021
\bibitem[Lamb et al.(2009)]{Lam08} Lamb, F.~K., Boutloukos, 
S., Van Wassenhove, S., Chamberlain, R.~T., Lo, K.~H., Clare, A., Yu, W., 
\& Miller, M.~C.\ 2009,
\apj, 706, 417
%
\bibitem[Leahy(2004)]{Leahy04} 
Leahy, D.~A.\ 2004, \apj, 613, 517 
\bibitem[Leahy et al.(2008)]{LMC08}
Leahy, D.~A., Morsink, S.~M., \& Cadeau, C.\ 2008, \apj, 672, 1119 
\bibitem[Leahy et al.(2009)]{LMCC09}
Leahy, D.~A., Morsink, S.~M., Chung, Y.-Y., \& Chou, Y.\ 2009, \apj, 691, 1235
\bibitem[Morsink et al.(2007)]{MLCB07}
Morsink, S. M., Leahy, D. A., Cadeau, C. \& Braga, J. 2007,
\apj, 663, 1244
\bibitem[Pandharipande(1971)]{eosA}
Pandharipande, V. R. 1971, Nucl. Phys. A, 174, 641
\bibitem[Pandharipande et al.(2006)]{eosL}
Pandharipande, V. R., Pines D., \& Smith, R. A. 1976, 
\apj, 208, 550
\bibitem[Papitto et al.(2005)]{Pap05}
Papitto, A., Menna, 
M.~T., Burderi, L., Di Salvo, T., D'Antona, F., \& Robba, N.~R.\ 2005, 
\apjl, 621, L113 
\bibitem[Papitto et al.(2009)]{Pap09} 
Papitto, A., Di Salvo, T., D'A{\`i}, A., Iaria, R., Burderi, L., 
Riggio, A., Menna, M.~T., \& Robba, N.~R.\ 2009, \aap, 493, L39 
\bibitem[Poutanen \& Gierli\'{n}ski(2003)]{PG03}
Poutanen, J. \& Gierli\'{n}ski, M. 2003, \mnras, 343, 1301
\bibitem[Poutanen(2006)]{P06} Poutanen, J.\ 2006, Advances 
in Space Research, 38, 2697
%
\bibitem[Sazonov \& Sunyaev(2001)]{SS01}
Sazonov, S. Y. \& Sunyaev, R. A. 2001, \aap, 373, 241
%
\bibitem[Wang et al.(2001)]{Wang} Wang, Z., et al.\ 2001, 
\apjl, 563, L61
\bibitem[Wang et al.(2009)]{Wang09} Wang, Z., Bassa, C., 
Cumming, A., \& Kaspi, V.~M.\ 2009, \apj, 694, 1115 
\bibitem[Wijnands \& van der Klis(1998)]{WvdK98} Wijnands, R., 
\& van der Klis, M.\ 1998, \nat, 394, 344
\bibitem[Wijnands(2006)]{W06} Wijnands, R.\ 2006, Trends in 
Pulsar Research, 53
\bibitem[Yakovlev \& Pethick(2004)]{YP04} Yakovlev, D.~G., \& Pethick, C.~J.\ 2004, \araa, 42, 169 




\end{thebibliography}
\end{document}